\newcommand\Msun{\hbox{M$_\odot$}}

\newcommand\eg{e.\,g.}

\newcommand\etal{et~al.}

\newcommand{\mnras}{MNRAS}
\newcommand{\procspie}{Proc.\ SPIE}
\newcommand{\aj}{AJ}
\newcommand{\repo}{https://bitbucket.org/iraklis\_k/starfish}

\documentclass[5p]{elsarticle}

\usepackage{lineno,hyperref}
\modulolinenumbers[250]

\journal{Astronomy \& Computing}

\biboptions{authoryear}

\begin{document}

\begin{frontmatter}

\title{The Starfish Diagram: Visualising Data Within the Context of Survey Samples.\tnoteref{mytitlenote}}
\tnotetext[mytitlenote]{Source code is available on \repo.}

\author[address1,address2]{Iraklis S. Konstantopoulos\fnref{myfootnote}}
\fntext[myfootnote]{John Stocker Postdoctoral Fellow.}

\address[address1]{Australian Astronomical Observatory, 
					PO Box 915
					North Ryde NSW 1670
					Australia; tel:~+61~(0)~2~9372~4859}
\address[address2]{ARC Centre of Excellence for All-sky Astrophysics (CAASTRO)}
\begin{abstract}
As astronomy becomes increasingly invested in large surveys, the ample representation of an individual target becomes a significant challenge. Tabulations of basic properties can convey the message in an absolute sense, but not within the context of the sample from which the individual is drawn. We present a novel but simple plot that simultaneously visualises the properties of the sample and the individual. Numbers and characters are kept at an absolute minimum to enable the stacking of such plots without introducing too much verbal information. Once the user becomes accustomed to their appearance, a set of \emph{starfish diagrams} provide a direct representation of the individual within a sample, or between various samples. The utility and versatility of the plot is demonstrated through its application to astrophysical data and sports statistics. We provide a brief description of the concept and the source code, which is simple to adapt to any statistical dataset, be it descriptive of physics, demographics, finance, and more. 
\end{abstract}

\begin{keyword}
{astronomical databases: miscellaneous}\sep 
{Visualization application domains: Scientific visualization}\sep 
{Visualization application domains: Visual analytics}\sep 
{Visualization application domains: Information visualization}\sep 
{Visualization: Visualization design and evaluation methods}\sep 
{Visualization theory, concepts and paradigms} 
\end{keyword}

\end{frontmatter}

\linenumbers

\section{Graphical Representation of Survey Samples\label{sec:intro}}
The future of Astrophysics lies with \emph{software telescopes}: facilities like the Square Kilometre Array \citep{ska} and Large Synoptic Survey Telescope \citep{lsst}, that will conduct large surveys with the aim of providing the community not with raw data, but `high-level' or derived science products. While the purpose of these facilities will be to establish statistically sound samples, the potential for studying individual targets is immense---especially considering the all-sky, all-wavelength coverage such surveys will  eventually accomplish. Take, for example, the UV-to-radio frequency Galaxy And Mass Assembly survey \citep[GAMA;][]{gama}; all studies emerging from this project use subsets of the full galaxy sample, however, the casual user may find this massive database more useful for the in-depth study of an individual galaxy, one that had not before been covered across a broad wavelength range. 

Current astronomical surveys, \eg\ the Sloan Digital Sky Survey \citep{sdss} and GAMA, typically offer rich sets of information through \emph{single-object viewers}, web-applications that combine images with graphical and tabulated information\footnote{http://skyserver.sdss3.org/dr9/en/tools/quicklook/quickobj.asp}$^,$\footnote{http://www.gama-survey.org/dr2/tools/sov.php}. Such applications can provide a quick and deep overview of a celestial object, however, they have not, so far, extended to pinpointing the place of these individual targets in the manifold parameter space covered by the sample-at-large. This context is crucial to interpreting the properties of the individual---it is not very useful to know that the stellar mass of a galaxy is $10^9~$\Msun\ if the user is not familiar with the mass function of galaxies in the universe. The requirement to convey such information is the purpose of the visualisation method presented in this paper. In Section~\ref{sec:concept} we present the concept and methodology, and in Section~\ref{sec:code} we go through some key segments of the source code. Section~\ref{sec:future} outlines the ongoing development of the code, primarily in the context of interactive web-applications, and evaluates some possibilities for future applications and further development.

\section{Conceptual Design\label{sec:concept}}
The visualisation method presented in this work was originally developed for the SAMI Galaxy Survey \citep[sami-survey.org;][]{sami}. The need for the simultaneous visualisation of individual and sample-wide properties emerged from work on the SAMI science archive, where the essence of a galaxy needs to be encapsulated within a single row of `quick-look' tabular information. The lack of context of listed values became immediately apparent: only an expert can readily interpret the meaning of a set of galaxy properties, and only in a general astrophysical context. In order to visualise the information, multiple properties need to be graphed and contextualised simultaneously, increasing the complexity of a plot. 

This realisation led to an initial conceptual design inspired partly from data visualisation in video games---the commonplace representation of \emph{player stats} as a set of spokes on a wheel. The maximum length of the spoke is 100\%, which requires no expertise to interpret. Translating this basic design to scientifically valuable information requires little adaptation. If we anchor the minimum and maximum values of a cross-sample measurable, say stellar mass, to the 0\% and 100\% of a full-radius spoke, we have clearly represented an individual measurement among the full range of values that exist in the whole sample. 

Unfortunately, however, the statistical distributions that underly many properties of physical systems (such as the galaxies studied by SAMI) are not described by a simple functional form---and rarely will two measured values be drawn from similar distributions. This way a simple `wagon wheel' design with its minimum-maximum indication does not convey the distribution of the measured properties; it does not aptly describe the statistics of the sample; and fails to place the individual in the context of the sample. 

Instead of plotting simple spokes we instead progressed to visualising the underlying statistics in their entirety as the frequency distribution of each value to be plotted. The resulting histogram was justified about the horizontal axis and the value of the target in question was indicated, The process was repeated for each plot variable, with each frequency distribution rotated about a pivot to simultaneously visualise all measurables. The spiky appearance owed to the irregular distribution of some histograms granted the plot its name: a \emph{starfish diagram}. 
\begin{figure*}[t]
\begin{center}
\includegraphics[width=0.4\textwidth]{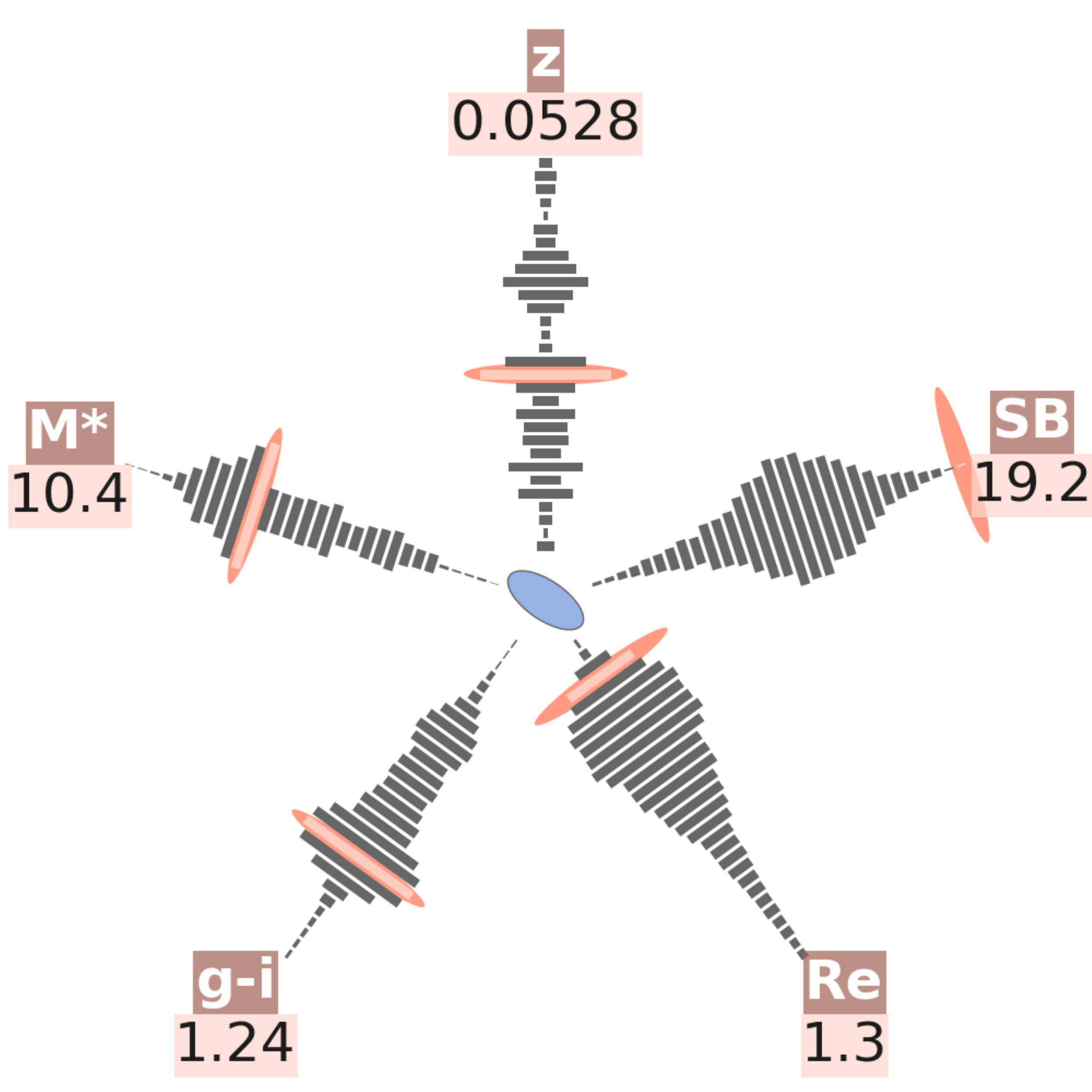}\hspace{20pt}
\includegraphics[width=0.4\textwidth]{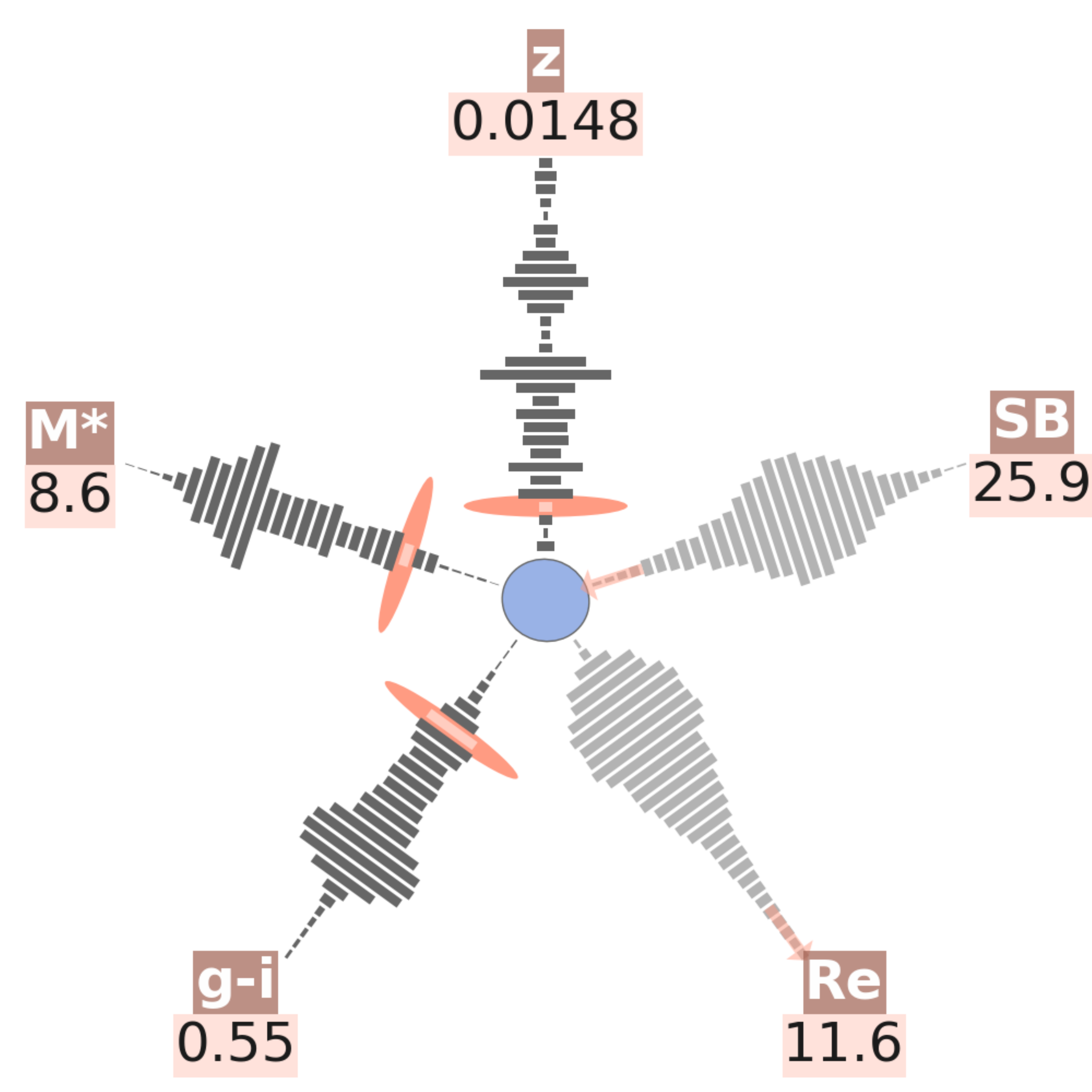}
\caption{A pair of starfish diagrams visualising information on two galaxies from the SAMI subsample of the GAMA survey (www.gama-survey.org/dr2). Each arm plots the frequency distribution of a sample-wide property, marked in shorthand as $z$ (redshift), $M^*$ (log stellar mass), $g-i$ (colour excess), $Re$ (effective radius), and $SB$ (surface brightness). The properties of the individual are marked as elliptical tick marks and in cases where these values are either not present or outside the 95\% plotted range, the starfish arm is rendered in a pale grey colour (but the values are still displayed). We avoid indicating any more information to maintain the visual simplicity, especially since the SAMI starfish are displayed as small cells in an HTML table. The central oval represents the ellipticity and position angle of the galaxy from our vantage point. While the plot could be simplified by displaying only a median value for each sample property, the varied source distributions render such simple statistics somewhat meaningless and ill-suited for direct comparison. For example, the median redshift falls on a large-scale structure `wall' in the GAMA fields, whereas surface brightness and effective radius are described by skewed Gaussian distributions. 
\label{fig:SAMI}}
\end{center}
\end{figure*}

The diagram in its default form displays five measured quantities, not just for anatomical correctness, but also to maintain its visual simplicity. More histograms can be drawn but care needs to be taken as some of the measured values may have bottom-heavy statistical distributions, and thus clutter the area near the pivot point. To circumvent this problem we added a void area about the pivot point, and used it to convey two extra pieces of information on the individual alone: the position angle of the galaxy and its ellipticity\footnote{Galaxies can be spheroidal or orbicular, or a combination of both. The inclination effect of disk-like galaxies seen off-axis changes their circularity, as perceived from our line of sight.}.

\section{Execution and Source Code\label{sec:code}}
Two starfish diagrams, produced with the code found on \repo, are shown in Figure~\ref{fig:SAMI}. Each arm is labelled with a shorthand title, so as not to clutter the display, along with the value of the individual at suitable precision. Each histogram contains the same number of bins as all variables are drawn from the same sample size, but differing sets of bins can be used without compromising the appearance of the diagram (this is set as a plot argument). 

Multiple starfish make for a very handy comparison tool: blinking through them one perceives the tick marks as moving along their axes and registers a quick mental contrast. One of the galaxies of Figure~\ref{fig:SAMI} has two values outside the plotted range (roughly limited to 95\% of all data), hence the respective histograms are plotted in a paler shade of gray to guide the eye away. The code offers the option of an annotation to indicate whether the value lies off the bottom or top of the plotted range (see Section~\ref{sec:future}). A similar course can be taken if there are missing values, including for the central oval that is simply not plotted if the information is not available. 

The code is written entirely in Python. We were committed to using common packages in order to simplify the adaptation of the code by others, hence the plot is created using \texttt{matplotlib}, the most widely used plotting package in Python (along with the associated \texttt{mpl\_toolkits} code). No specialist packages are used: arrays are stored using \texttt{numpy}, the standard numerics package, and data are stored using the \texttt{pickle} package (a Python-native binary data format), which completes the set of codes imported by the plot code in the default mode. The default input is tables of comma-separated values (CSV), although the user can opt to import tables in the HDF5 format\footnote{http://www.hdfgroup.org/HDF5/}. This feature is secondary in the general \texttt{starfish} package, but the primary mode of data ingestions in the original code, which was developed specifically for the SAMI data archive (Konstantopoulos~\etal, in preparation for submission in this Journal).  

The data preparation code, \texttt{frequency()}, scans the input file and determines its type, either CSV or HDF5. It requires the user to define a range of acceptable variables for each array to be plotted, and values outside this range are masked, rather than discarded, in order to maintain the indexing and dimensionality of all arrays (and enable queries across a homogeneous sample). We store all data in \texttt{numpy} masked arrays. In the generalised version of the code these values are supplied by the user as input arguments, but this segment of code is best explained through the following SAMI prototype code segment\footnote{Available on https://bitbucket.org/iraklis\_k/sami-database/}: 
\begin{small}
\begin{verbatim}
# Name useful variables. 
names = ['z_spec', 'Mstar', 'g-i', 'r_e', 'mu_re', 
         'ellip', 'PA', 'OBS_SAMI', 'CATID']

# Make list of masked variable arrays. 
myVariables = [
    ma.masked_less(tab[names[0]].data, 0.0),
    ma.masked_greater(tab[names[1]].data, 12.),
    ma.masked_outside(tab[names[2]].data, 0., 1.8),
    ma.masked_outside(tab[names[3]].data, 0., 10.),
    ma.masked_outside(tab[names[4]].data, 19., 25.),
    ma.masked_outside(tab[names[5]].data, 0., 90.), 
    ma.masked_outside(tab[names[6]].data, 0., 180.), 
    tab[names[7]].data, 
    tab[names[8]].data]
\end{verbatim}
\end{small}

These arrays are included in a Python list, along with the galaxy ID and a boolean flag (revealing whether the galaxy has been observed) as regular arrays to avoid the overhead of masking, as we know them have valid values throughout. In the generalised case these are also masked, but since this code is designed for survey operations, it is optimised to a high level to avoid needless computation. 

Once the list of masked arrays has been created it is either \emph{pickled} as a dictionary, or returned to another code:
\begin{small}
\begin{verbatim}
if pickle:
    # Create and pickle a dictionary called data. 
    data = {'Name' : names, 
            'Bins' : binsList, 
            'Frequency' : frequencyList, 
            'VarArrays' : myVariables}

    fpickle = open(fname, 'wb')
    pickle.dump(data, fpickle)
    fpickle.close()
else:
    return(myPlotVariables, binsList, 
    	   frequencyList, VarArrays)
\end{verbatim}
\end{small}
\ldots where \texttt{fname} is the user-supplied pickle file name. The \texttt{plot()} code then unjars the pickled data file and only deals with visualising the information is contains:
%

The code is modularised to comply with survey operations, as there is no reason for the frequency distributions to be calculated every time a starfish is to be created. Pickling also ensures a record of what information has gone into the creation of a plot, which provides a suitable quality control file. The plot code loops over the number of named variables, also read from the pickled file, to read the frequency distributions once, before looping over all galaxies to be plotted. The pickle file also contains variable names so that no manual input is required by \texttt{plot()} apart from the pickle filename. 

The first action is to resample the array to a canvas with an appropriate coordinate system, from the edge of the area allocated to the central oval to the edge of the \texttt{matplotlib} window: 
\begin{small}
\begin{verbatim}
# Resample variables array to fit canvas. 
axmin = 0.5 ; axmax = 1.0; axscrunch = 0.88
def resample(var):
    a = var-min(var)
    a = a/max(a)
    a = a/(axmax/(axscrunch*origins[1]-bodyWidth)) +\
         (axmin+bodyWidth) 
    return(a)
\end{verbatim}
\end{small}
\ldots where \texttt{bodyWidth}, the size of the central shape, is user-defined. Then an axis-free canvas must be set up: 
\begin{small}
\begin{verbatim}
# Set up a white canvas with a Cartesian grid. 
fig = plt.figure(figsize=(10, 10), dpi=100)
ax = fig.add_axes([0, 0, 1, 1])
plt.axis('off') ; ax.set_aspect('equal')
cartesian = ax.transData
# Set the coord transformation centre of reference.
origins = [0.5, 0.45]
ctrOfRotation = cartesian.transform(origins)
\end{verbatim}
\end{small}

At this stage the loop over all named variables can commence: the code reads the histogram ingredients from the pickle; determines whether the value is masked or uncovered and defined the value accordingly; normalises the frequency amplitude to user-defined canvas units; resamples the array using the function defined above (outside the loop); and defines \texttt{t}, the coordinate transformation matrix. Looping over index \texttt{[i]}:
\begin{small}
\begin{verbatim}
# Read pickled histogram data.
freqVariable = myData['Frequency'][i]
ticksVariable = myData['Bins'][i]
    
if len(myData['VarArrays'][i].mask.ravel())==1:
    """ Then no values masked. """
    mask = False
    myValue = myData['VarArrays'][i][index]
else:
    masked = myData['VarArrays'][i].mask
    if masked[index]:
        mask = True
        myValue =\
            myData['VarArrays'][i].data[index]
    else:
        mask = False
        myValue =\
            myData['VarArrays'][i][index]

# Normalise frequency axis to X canvas units. 
normVariable =\ 
    maxBarHeight*freqVariable/max(freqVariable)
    
# Resample the array to fit the canvas. 
starfishArm = resample(ticksVariable)
    
# Define cordinate transformation (x, y, angle).
transform = Affine2D().rotate_deg_around(
    ctrOfRotation[0], ctrOfRotation[1], angles[i]+90)
t = cartesian + transform
\end{verbatim}
\end{small}

Once that is all complete, the bin where the value of the individual galaxy belongs is identified through the index array, which is why masked arrays are employed in the first place: 
\begin{small}
\begin{verbatim}
# Identify the bin to which value corresponds. 
if not mask:
    whereTick =\ 
        [np.abs(ticksVariable-myValue).argmin()]
    tickMark = starfishArm[whereTick]
\end{verbatim}
\end{small}
\ldots before continuing on to standard \texttt{matplotlib} to arrange the plot: a marker (ellipse-\emph{patch}), histogram bars (rectangle patches) and dots in place of empty bins, and text annotations for values and variable names.

The code can currently only save the plot as a PNG file, owing to limitations in the bounding box of the \texttt{matplotlib} canvas. This cannot at the moment be changed simply to print to PDF, encapsulated postscript or scalable vector graphics. In the following section we will describe current work to implement scalable graphics printing and other functionality that is planned for the near future. 

\section{Further Development\label{sec:future}}
The starfish diagram is being developed beyond the prototype presented in this article. Since it is meant for the web, the next iteration will be printed as a scalable vector graphics item, rather than a static bitmap. In that way it will achieve its potential of interactivity. By assigning an \emph{HTML tooltip} to each plot item (each \texttt{matplotlib} patch), the user will be able to access a wealth of information. This interaction will be accomplished by a series of tools, primarily a \emph{hover} tool that reveals information when the cursor hovers above a patch. Connecting various plots, so that actions on one plot reflect on all others, will also add functionality and increase the capability for data exploration. 

The starfish diagram is meant for server-based, client-side data exploration, so the extensions we are currently developing are oriented toward the use of JavaScript to embed all this information into the HTML source. The compactness of JavaScript Object Notation is very valuable in reducing the amount of code that the server would have to transfer in indent-specific languages, such as Python. And if the data are embedded in the HTML verbiage, then the onus falls on the browser to process all interaction between the scientist and the plot. By making all plot updates \emph{asynchronous}\footnote{http://www.adaptivepath.com/ideas/ajax-new-approach-web-applications/}, client-server communication is minimised and data can be explored without the necessity of a fast internet connection. This is our vision for the future of the code, and all development will be reflected on the starfish repository. 

\subsection{Applications Outside Astrophysics\label{sec:future}}
\begin{figure*}[htbp]
\begin{center}
\includegraphics[width=0.4\textwidth]{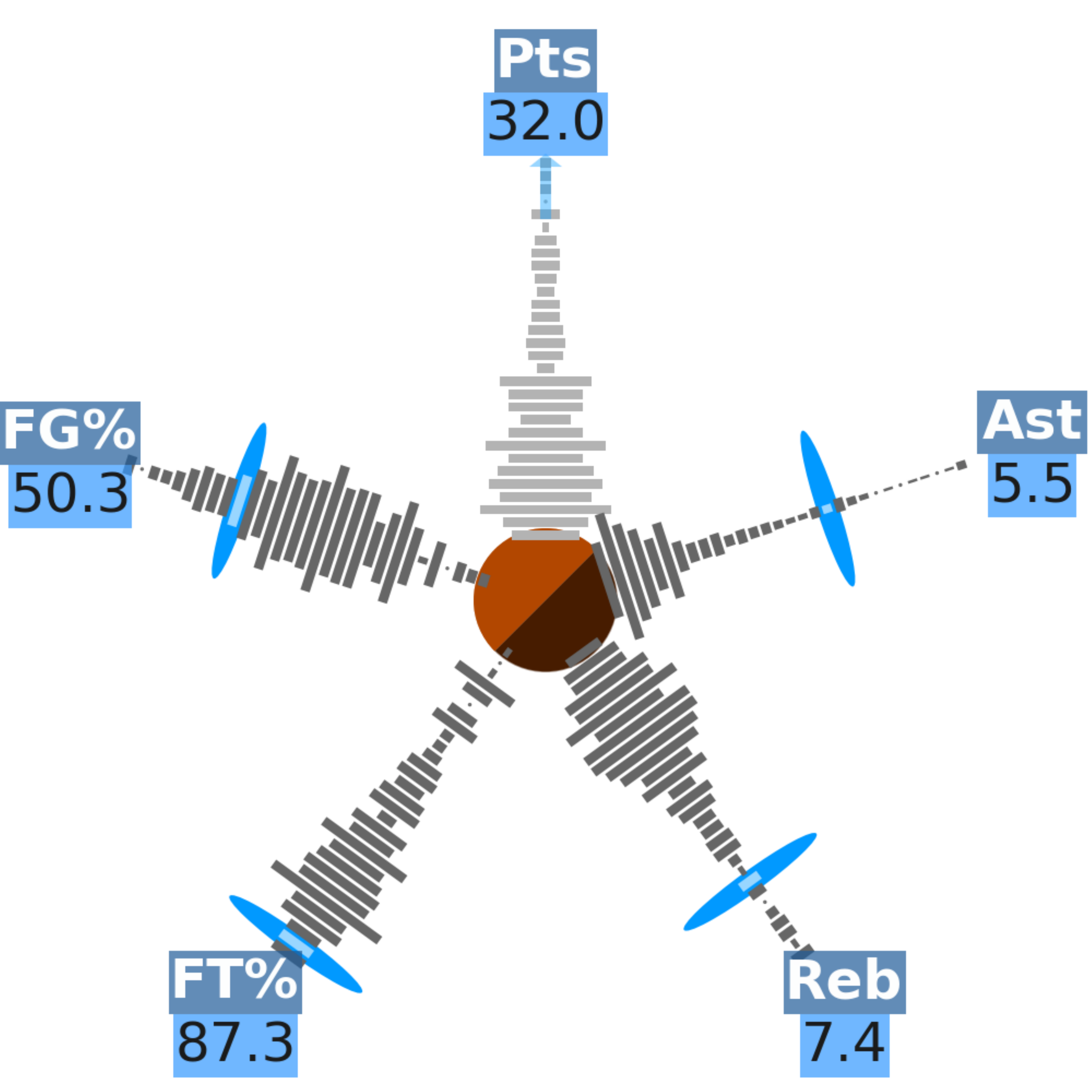}
\includegraphics[width=0.4\textwidth]{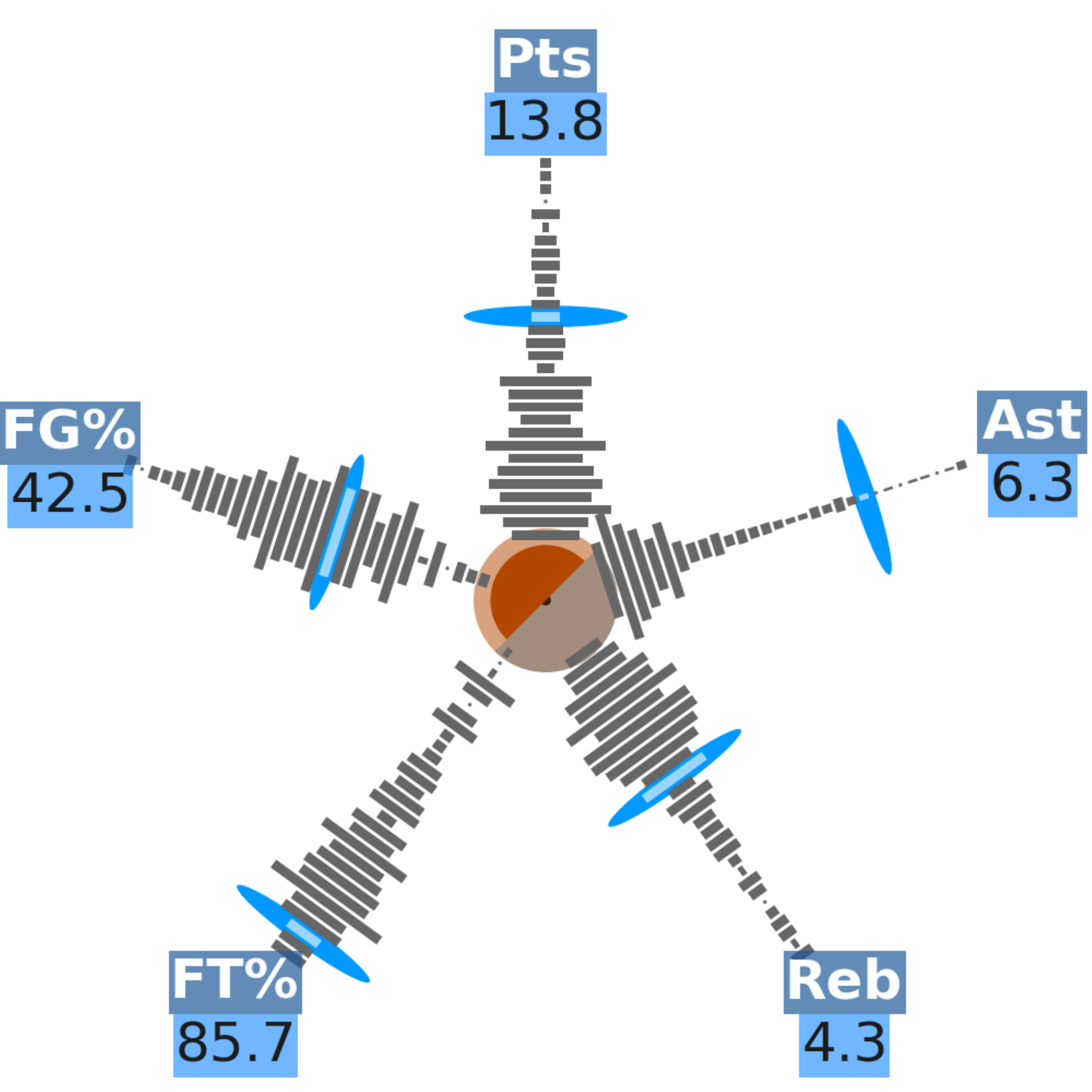}
\caption{An application on sporting statistics, created with the generalised version of the code available online. The starfish on the left shows five key performance indicators of basketball players Kevin Durant, voted `most valuable player' in the US league's 2013-14 regular season. Durant's point scoring (Pts) is off the top of the displayed range, as indicated by the blue arrow, and his overall performance appears to justify his MVP award (counterclockwise: field goal and free throw success rates, rebounds, assists). The central shape in this case counts minutes played per game (orange semicircle) and games played (brown semicircle), normalised to the respective maxima. In order to demonstrate transparency in these semicircles, we show the starfish diagram for Kobe Bryant, who was injured for most of the season: his brown circle is mostly grey, corresponding to the six games he played of a possible 84. For aesthetic effect the circles have been oversized as compared to the galaxy starfish of Figure~\ref{fig:SAMI}, while the maximum height of the histogram bars has been decreased using an input argument. Note also that empty bins are marked with dots.}\label{fig:bball}
\end{center}
\end{figure*}

This diagram is essentially a statistical tool and as such its use is not restricted to describing physical systems. In Figure~\ref{fig:bball} we show an application of the diagram in visualising the performance statistics of basketball player Kevin Durant, placed in the context of his colleagues in the Western Division of the National Basketball Association (USA) during the 2013-14 season\footnote{Statistics obtained from http://stats.nba.com under fair use.}. Durant was voted `most valuable player' during that season and his starfish diagram reflects that amply, displaying exceptional performance in five key indicators. His scoring is off the top of the displayed range, indicated by a subtly plotted arrow. The central shape has been generalised, since athletes are not galaxy-like, to a pair of semicircles representing more statistics. In this case, they plot the player's participation in terms of the number of games (brown) and minutes played per game (orange). To reinforce the utility of the starfish diagram as a comparison tool, we contrast the performance of Durant with basketball superstar Kobe Bryant, who spent most of the season injured and unable to join his team---hence the mostly transparent brown semicircle. This version of the diagram is created with the generalised code available online. 

\section{Summary\label{sec:summary}}
We have presented the \emph{starfish diagram}, a novel method to simultaneously visualise the statistical properties of individual items within large samples and the properties of the sample as a whole. This is designed for large surveys where an individual can be placed in the context of the general population. It was developed for an astrophysical survey, but the range of applications for the starfish diagram is far broader, including, but not limited to, the demographics of individuals or subsamples drawn from census data; the properties of particles in accelerator experiments; socioeconomic metrics for nations in the world stage; and going full circle to the inspiration of this plot, to the statistics of athletes' performance. The code is freely available online (including an example function to help the user become acquainted with the code) and we hope this will help researchers from a variety of fields visualise their information in a direct, colourful, accessible, and concise manner. 

\section*{Acknowledgements}
I thank Amanda E.\ Bauer and Aaron S.\ G.\ Robotham for suggestions on the plot and manuscript. 

ISK is the recipient of a John Stocker Postdoctoral Fellowship from the Science and Industry Endowment Fund. 

GAMA is a joint European-Australasian project based around a spectroscopic campaign using the Anglo-Australian Telescope. The GAMA input catalogue is based on data taken from the Sloan Digital Sky Survey and the UKIRT Infrared Deep Sky Survey. Complementary imaging of the GAMA regions is being obtained by a number of independent survey programs including GALEX MIS, VST KiDS, VISTA VIKING, WISE, Herschel-ATLAS, GMRT and ASKAP providing UV to radio coverage. GAMA is funded by the STFC (UK), the ARC (Australia), the AAO, and the participating institutions. The GAMA website is http://www.gama-survey.org/ .

The SAMI Galaxy Survey is based on observation made at the Anglo-Australian Telescope. The Sydney-AAO Multi-object Integral field spectrograph (SAMI) was developed jointly by the University of Sydney and the Australian Astronomical Observatory. The SAMI input catalogue is based on data taken from the Sloan Digital Sky Survey, the GAMA Survey and the VST ATLAS Survey. The SAMI Galaxy Survey is funded by the Australian Research Council Centre of Excellence for All-sky Astrophysics (CAASTRO), through project number CE110001020, and other participating institutions. The SAMI Galaxy Survey website is http://sami-survey.org/ .


\end{document}